\documentclass
	[
    a4paper,twocolumn,twoside,12pt,showkeys, 
    ]
	{revtex4}
\usepackage{amsmath,amssymb}	% AMS math fonts
\usepackage{bm}					% bold math
\usepackage{color}					% for color text and text box
\usepackage{graphicx}				% Include figure files
\usepackage{dcolumn}				% Align table columns on decimal point
\usepackage[
  colorlinks=true, citebordercolor={0 0 0}, linkcolor=blue, citecolor=blue, urlcolor=blue
  ]{hyperref}
\usepackage{setspace}

\makeatletter
\def\frontmatter@abstractwidth{0.9\textwidth}	% abstract width
\makeatother

\usepackage{siunitx}
\usepackage{ulem}

\begin{document}
%\setcounter{page}{\value{startPage}}

%%%%% DO NOT CHANGE THE ABOVE HERADERS %%%%%

%%%%% Several usefull commands %%%%%
%%%%% YOU CAN ADD YOUR OWN COMMANDS HERE %%%%%
\newcommand{\By}{$\times$}
\newcommand{\SqrtBy}[2]{$\sqrt{#1}$\kern0.2ex$\times$\kern-0.2ex$\sqrt{#2}$}
\newcommand{\Degree}{$^\circ$}
\newcommand{\DegreeC}{$^\circ$C\,}
\newcommand{\Ohmcm}{$\Omega\cdot$cm}
\newcommand{\kay}{cm^{-1}}
\newcommand{\Sp}[1]{$\mathrm{sp}^{#1}$}
\newcommand{\EF}{$E_{\mathrm{F}}$}

\title{%%%%% PUT TITLE OF THE PAPER HERE %%%%%
Unconventional thermal conductivity of suspended zigzag graphene nanomesh
}

%%%%% PUT AUTHOR INFORMATIONS HERE %%%%%

\author{Takamoto Yokosawa$^1$}
\author{Tomohiro Matsui$^1$}
	\email[Corresponding author: ]{Tomohiro.Matsui@anritsu.com}

\affiliation{%
$^1$Advanced Research Laboratory, Anritsu corporation, 5-1-1 Onna, Atsugi, Kanagawa 243-8555, Japan
}

%\date{\today}

\begin{abstract}
\vspace*{1mm}

Compared to the study of graphene itself, the study of nano-structured graphene is rather limited because it is difficult to prepare atomically ordered edges.
In this study, we have fabricated a periodically patterned mesh structure of graphene with atomically precise zigzag edges (zGNM: zigzag graphene nanomesh) and studied its thermal conductivity ($\kappa$) by opto-thermal Raman measurement.
Unintuitively, it is found that the $\kappa$ of zGNM of $2,3$~monolayers (MLs) thick is inversely proportional to the nanoribbon width ($W$), while that of zGNM of 5$\sim$10~MLs thick is independent of $W$ down to 30~nm.
Since the $\kappa$ of suspended zigzag graphene nanoribbons (zGNRs) is suppressed by decreasing $W$, this nonclassical behavior of zGNM is due to the mesh structure.
In addition, zGNRs show a higher $\kappa$ than GNRs with atomically rough edges.
This is probably due to the atomically ordered zigzag edges.
\end{abstract}

\keywords{%%%%% PUT KEYWORDS HERE %%%%%
graphene, thermal conductivity, zigzag edge, nanomesh, nanoribbon
}

%%%%%%%%%% DO NOT CHANGE THE FOLLOWING %%%%%%%%%%
%\date[]{Received \receivedData; Accepted \acceptedData; Published \pulishedData} 
\maketitle
%\thispagestyle{ejssntFP}
%\pagestyle{ejssnt}
%%%%%%%%%% DO NOT CHANGE THE ABOVE %%%%%%%%%%
\newpage

%%%%%%%%%%%%%%%%%%%%%%
%% BODY OF THE PAPER
%%%%%%%%%%%%%%%%%%%%%%

%%%%%%%%%%%%%%%%%%%%%%
\section{Introduction}
%%%%%%%%%%%%%%%%%%%%%%

Graphene shows unique electronic \cite{Novoselov2004, Novoselov2005, Zhang2005, Neto2009}, thermal \cite{Balandin2008, Pop2012, Fu2019}, and mechanical \cite{Frank2007, Lee2008, Papageorgiu2017} properties from both fundamental and applied research perspectives.
Nano-structuring can extend these features or create even more unique features because a wave-like nature of quasi-particles emerges when the length scale of a device becomes shorter than the mean free path of the quasi-particles.
However, study and application of nano-fabricated graphene is limited because control of graphene edges is difficult.
The edge property dominates the performance of nano-devices, since the edge ratio in the whole surface area increases as the device scale decreases.
Especially in the case of graphene, the edge property is significantly affected by the edge atomic structure.
For example, the zigzag (zz) edge possesses a flat band at the Dirac point energy, which results in an electronic state localized around the edge since the sublattice symmetry is broken around a zz edge \cite{Fujita1996, NiimiMatsuiKambaraEtAl2005, KobayashiFukuiEnokiEtAl2005, NiimiMatsuiKambaraEtAl2006}.
The edge state can even be spin-polarized for graphene nanoribbons (GNRs) terminated by zz edges (zGNRs).

On the other hand, reflecting the increasing demand for thermal management, study of the thermal properties of nano-structured graphene and the effect of edges is growing in importance.
M-H. Bae $et$ $al.$ \cite{Bae2013} showed that the thermal conductivity ($\kappa$) of GNRs supported on an SiO$_2$ substrate, which were fabricated by oxygen plasma etching of exfoliated graphene, decreases by decreasing the nanoribbon width ($W$). 
In this study, the $\kappa$ of the graphene device was obtained by subtracting the $\kappa$ of the substrate without graphene from the $\kappa$ of the whole device.
Similarly, a study of graphene nanomesh (GNM) suspended from substrate showed that $\kappa$ is decreased by decreasing the mesh neck width \cite{Oh2017}. 
In this case, the GNM was fabricated by reactive ion etching of CVD graphene.
Note that, for both cases, $\kappa$ is suppressed by nano-structuring as can be expected classically.
However, these studies are insufficient for discussing the true effect of nano-structuring because the edges of these devices must be rough in atomic scale. 
In addition, the nano-holes of the GNM \cite{Oh2017} are irregularly shaped and they are positioned irregularly.

Contrary to intuition, enhancement of thermal conductance ($K$) by nano-structuring is observed for suspended silicon nitride (SiN) membranes with periodically aligned circular holes at low temperatures below 1 K \cite{Zen2014}. 
They showed that a two-dimensional phononic crystal (PnC) with a smaller period has a higher $K$ than that with a larger period.
Such enhancement of $K$ can be understood as an appearance of the wave nature of phonons, which becomes prominent when the mean free path of the phonon becomes longer than the system size by nano-structuring at low temperatures \cite{Anufriev2015}. 

Recently, we have established a technique to fabricate graphene at nanometer scale, with zz edges \cite{Yokosawa2022}. 
The nano-structures are based on hexagonal nanopits, the edges of which are confirmed not only to be atomically precise zz structure \cite{Matsui2019, Amend2018}, but also whose edge C atoms are terminated by only one H atom \cite{Ochi2023}.
Using this technique, a hexagonal network of zGNR can be obtained by configuring hexagonal nanopits in a triangular lattice. 
This structure can be considered as an antidot lattice or a PnC depending on the physics of concern. 
Here, we call it zigzag GNM (zGNM).
A zGNR and its array can also be obtained in between elongated hexagonal nanopits with this technique. 
One can also suspend these nano-structures from an SiO$_2$ substrate by attaching electrodes and etching away the substrate.

In this study, the $\kappa$ of the suspended zGNM and zGNR were studied and it was found that the $\kappa$ of zGNM was not suppressed by decreasing $W$.
The $\kappa$ of the zGNM of less than three monolayers (MLs) thick was inversely proportional to $W$, while that of 5$\sim$10~MLs thick was independent of $W$.
On the other hand, the $\kappa$ of zGNR was decreased by decreasing $W$. 
Therefore, the unintuitive behavior of zGNM is related to the mesh structure.
At the same time, the comparison between the zGNR and GNR in \cite{Bae2013} suggests that $\kappa$ can be higher when the edge is a zz structure.

%%%%%%%%%%%%%%%%%%%%%%%%%%%%%%%%%%%%%%%%%%%%%%%%%%%%%%%%%%%%
\begin{figure}[b]
%\centering
\includegraphics[width=1.0\linewidth]{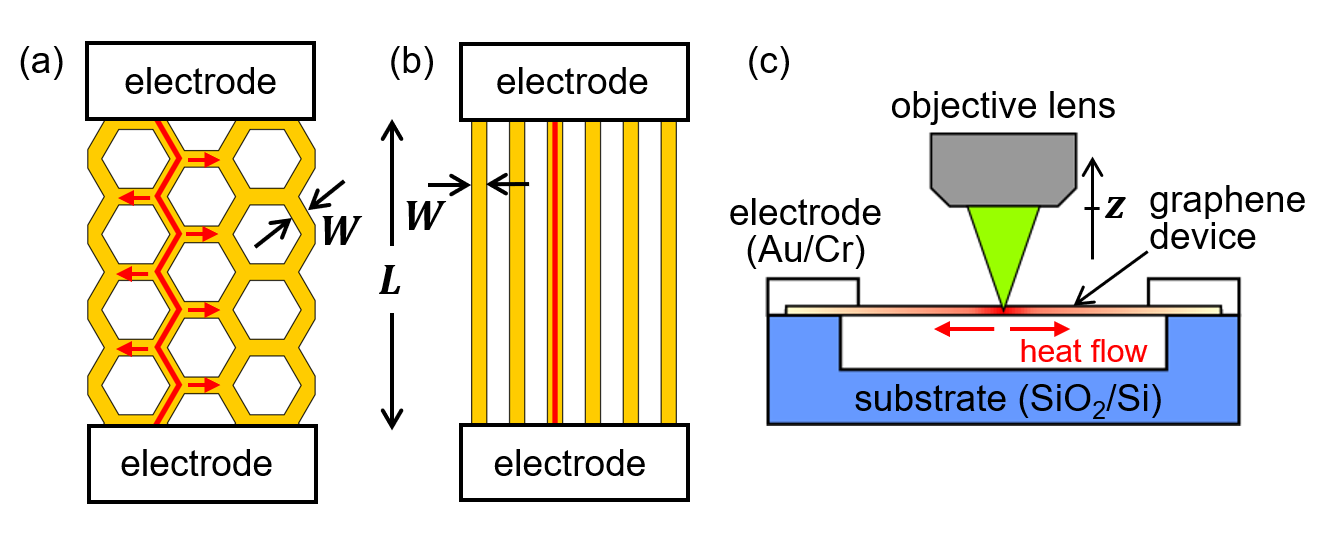}
\caption{
Schematic diagram of (a) zGNM and (b) zGNR devices and (c) the experimental configuration.
The nanoribbon width ($W$), device length ($L$) and objective lens height difference ($z$) were defined as illustrated.
The red zigzag line in (a) and red straight line in (b) are the assumed quasi-1D and 1D heat paths, respectively.
}
\label{fig:configuration}
\end{figure}
%%%%%%%%%%%%%%%%%%%%%%%%%%%%%%%%%%%%%%%%%%%%%%%%%%%%%%%%%%%%%

%%%%%%%%%%%%%%%%%%%%%%%%%%%%%%%%%
\section{Experimental}
%%%%%%%%%%%%%%%%%%%%%%%%%%%%%%%%%

The zGNM was fabricated from exfoliated graphene of 2 to 10 MLs thick.
The graphene thickness ($t$) was determined with a laser Raman microscope (inVia Raman microscope, Renishaw plc.) and a peak-force tapping-mode atomic force microscope (AFM, Dimension XR, Bruker).
The length of the device was set as 2 $\mathrm{\mu}$m.
The $W$ was controlled between 20 and 160~nm by changing the nanopit distance, while keeping the nanopit size as $346\pm27$~nm. 
The periodicity of the hexagonal nanopits was 350 to 540 nm for zGNM.
$\kappa$ was obtained by opto-thermal Raman measurement \cite{Judek2015, Ghosh2010, Cai2010, Balandin2008} using the laser Raman microscope with a laser wavelength of 532~nm and a $\times100$ objective lens.
The laser spot size is estimated to be about 680~nm.
The Raman laser, which acts as a heat source, is irradiated at the central part of the device.
On the other hand, electrodes (Au/Cr), which were attached to the devices as shown in fig.~\ref{fig:configuration}(a)(b), act as heat baths at room temperature.
Therefore, the heat flows from the central part of the device toward the electrodes as illustrated in fig.~\ref{fig:configuration}(c).

This measurement technique is based on the feature that the G-band of graphene shifts by temperature ($T$).
Namely, the temperature change ($\Delta T$) can be obtained from the G-band frequency shift ($\Delta\omega_\mathrm{G}$) as shown below.
\begin{equation}
  \Delta T = \frac{1}{\chi}\Delta\omega_\mathrm{G}.
  \label{eq:Gband-T}
\end{equation}
The coefficient $\chi = -1.6 \times 10^{-2}$ (cm$^{-1}$/K) \cite{Calizo2007,Judek2015} was used in this study.
On the other hand, the amount of heat absorbed by graphene ($P_\mathrm{abs}$) is controlled by the excitation laser power ($P$), the absorptivity of the monolayer graphene ($\alpha_{\mathrm{ML}}\sim2.3$~\%), the number of graphene layers ($N$), and the ratio of the graphene area occupying the laser spot ($A$) as follows.
\begin{equation}
  P_\mathrm{abs} = P(1-\mathrm{exp}(-\alpha_{\mathrm{ML}}N))\cdot A.
  \label{P_abs}
\end{equation}

%%%%%%%%%%%%%%%%%%%%%%%%%%%%%%%%%%%%%%%%%%%%%%%%%%%%%%%%%%%%
\begin{figure}[t]
%\centering
\includegraphics[width=1.0\linewidth]{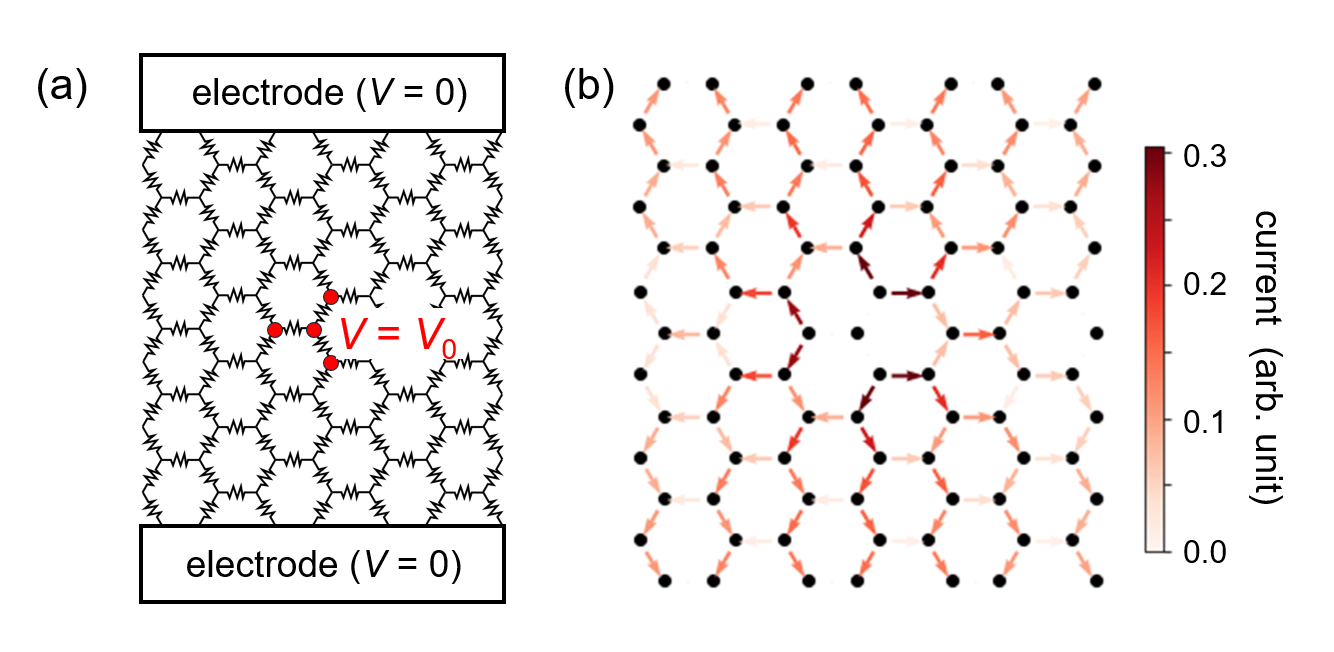}
\caption{
(a) Schematic diagram of a hexagonal network of resistance to simulate electronic current that flows in the mesh structure, and (b) simulated current map.
Potential $V$ at the position, which corresponds to the laser-irradiated area, is set as $V_0$, while $V$ is set as $0$ at electrodes. 
}
\label{fig:current model}
\end{figure}
%%%%%%%%%%%%%%%%%%%%%%%%%%%%%%%%%%%%%%%%%%%%%%%%%%%%%%%%%%%%%

Since thermal conductance ($K$) is the coefficient of $\Delta T$ when heat is applied to a substance, i.e., $P_\mathrm{abs}=K\Delta T$, $K$ can be described using $P_\mathrm{abs}$ and $\omega_\mathrm{G}$ as shown below.
\begin{equation}
  K = \chi\left(\frac{\partial\omega_\mathrm{G}}{\partial P_\mathrm{abs}}\right)^{-1}.
  \label{eq:K}
\end{equation}
Although $\kappa$ can be obtained from $K$ and the device dimensions, the definition of the device dimension is complicated for a zGNM different from the case for a ribbon.
In this analysis, we first assume a quasi-1D heat path as shown by the red zigzag line in fig.~\ref{fig:configuration}(a).
However, with this assumption, the $\kappa$ is over-estimated because heat can also flow in the direction perpendicular to the quasi-1D path (red arrows in fig.~\ref{fig:configuration}(a)). 
Therefore, coefficient $a$ was introduced to compensate for the over-estimation as follows using the length of the quasi-1D path ($L_\mathrm{eff}$).
\begin{equation}
  \kappa = \frac{L_\mathrm{eff}/2}{tW}K\times\frac{1}{a}
  \label{eq:kappa-K}
\end{equation}

Coefficient $a$ was estimated from numerical simulations of electronic current, which flows in the resistance network shown in fig.~\ref{fig:current model}(a).
Here, the electric potential ($V$) of two electrodes was set as $0$, while $V$ at the position corresponding to the Raman laser spot was set as $V_0$.
The current can be obtained by simultaneously combining Kirchhoff's equation.
Figure~\ref{fig:current model}(b) shows a map of the simulated current flowing in the circuit.
Coefficient $a$ was then estimated from the current flowing out from the $V = V_0$ points.
Although $a$ was different depending on the position of $V = V_0$ and the circuit size, it was estimated to be between 1.8 and 2.0 in our device dimensions. 
Therefore, $a$ was set as 1.9 in this study.

%%%%%%%%%%%%%%%%%%%%%%%%%%%%%%%%%%%%%%%%%%%%%%%%%%%%%%%%%%%%
\begin{figure}[t]
%\centering
\includegraphics[width=1.0\linewidth]{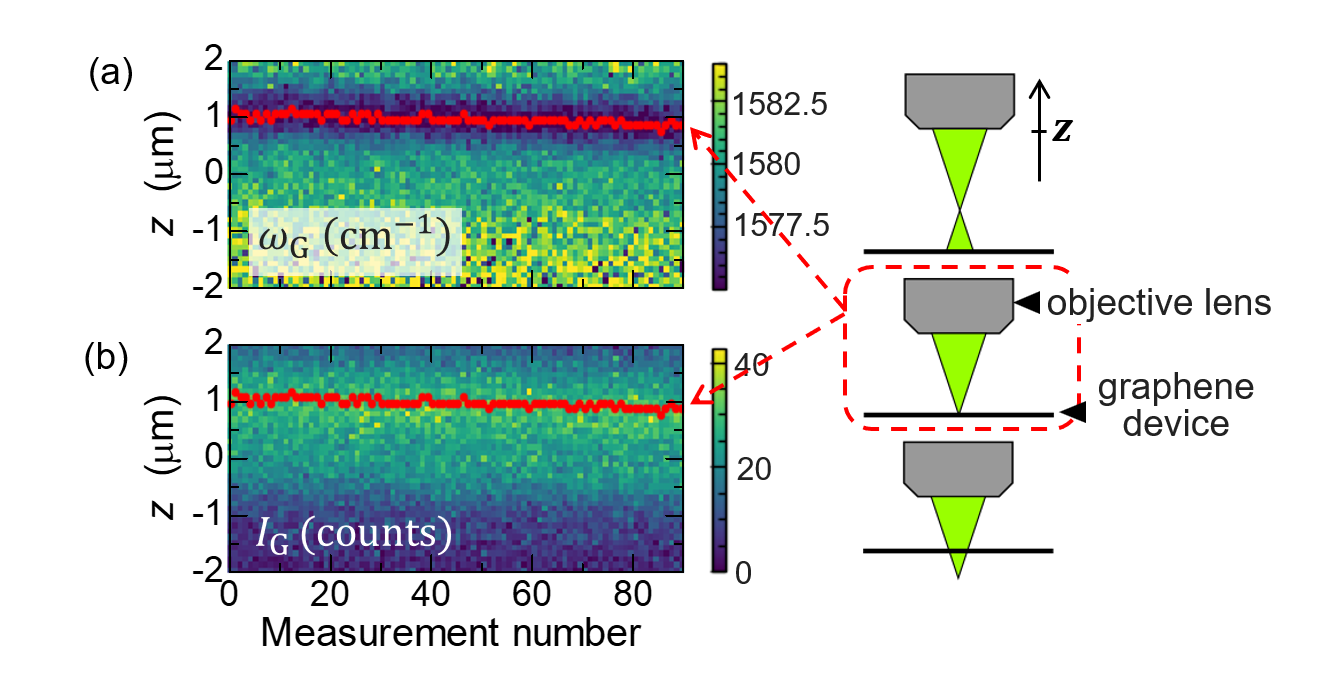}
\caption{
The (a) $\omega_\mathrm{G}$ and (b) $I_\mathrm{G}$ obtained by changing $z$ during 90 successive measurements for a zGNM device at $P_\mathrm{abs}=0.19$~mW.
The right-hand figures illustrate the configuration of the objective lens and graphene device. 
The $\omega_\mathrm{G}$ becomes minimum, while the $I_\mathrm{G}$ becomes maximum, where the laser through the objective lens focuses at the sample position.
The focused $z$ changes gradually due to thermal drift.
}
\label{fig:drift}
\end{figure}
%%%%%%%%%%%%%%%%%%%%%%%%%%%%%%%%%%%%%%%%%%%%%%%%%%%%%%%%%%%%%

For the opto-thermal Raman measurement, two additional processes were undertaken to obtain reliable data.
First, the measurement was performed in short time ($\sim$0.02~s) to prevent degradation of suspended graphene by continuous laser irradiation, and the measurement was repeated many times to earn a sufficient signal-to-noise ratio.
It was confirmed that the $\omega_\mathrm{G0}$, when the $\omega_\mathrm{G}$ was obtained at a low enough $P$, was $1583\pm2$~cm$^{-1}$ for all devices both before and after the measurement, which indicates that the device was not hung down between electrodes and no unexpected strain was applied to the devices.
Second, each spectrum was obtained by changing the height of the objective lens ($z$) to overcome the focus misalignment due to thermal drift during the whole measurement.
Figure~\ref{fig:drift} shows the $\omega_\mathrm{G}$ and $I_\mathrm{G}$ depending on $z$ during successive measurements.
They demonstrate that the height, where $\omega_\mathrm{G}$ becomes minimum and $I_\mathrm{G}$ becomes maximum, changes gradually during the measurement. 
The minimum values of $\omega_\mathrm{G}$ in each measurement were adopted and averaged in this study.

%%%%%%%%%%%%%%%%%%%%%%%%%%%%%%%%%
\section{Results and Discussion}
%%%%%%%%%%%%%%%%%%%%%%%%%%%%%%%%%

%%%%%%%%%%%%%%%%%%%%%%%%%%%%%%%%%%%%%%%%%%%%%%%%%%%%%%%%%%%%
\begin{figure}[t]
%\centering
\includegraphics[width=1.0\linewidth]{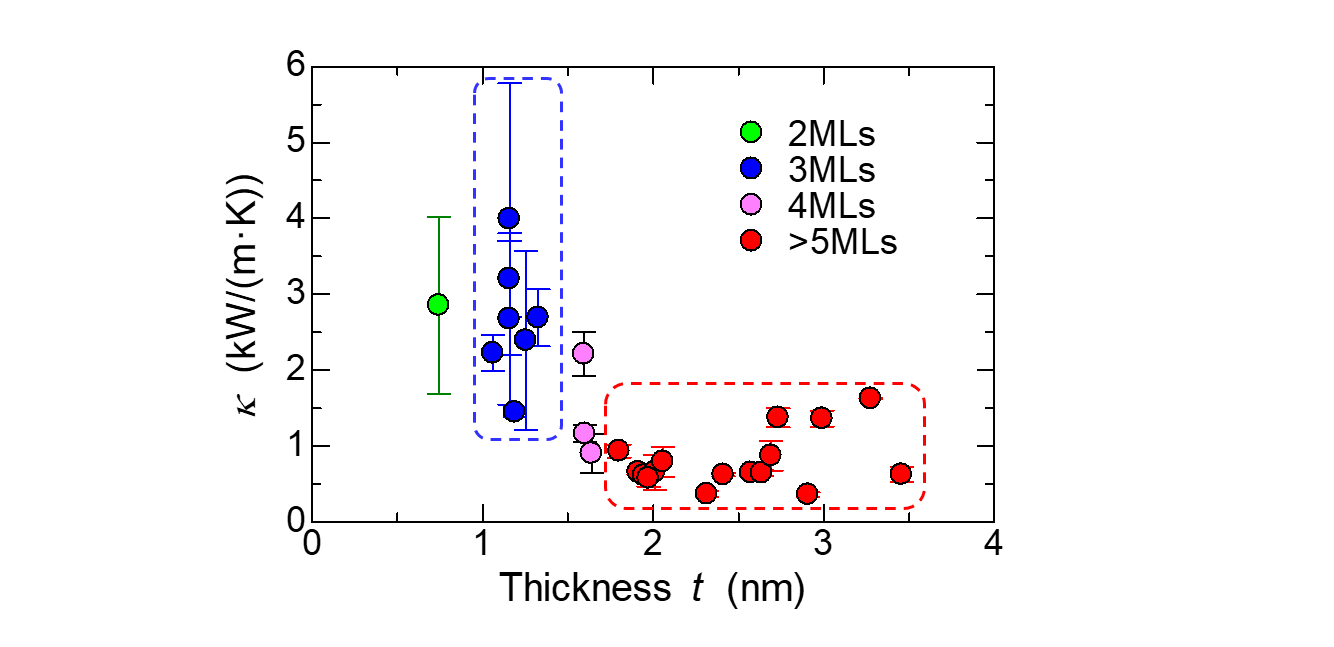}
\caption{
The $t$ dependence of the $\kappa$ of suspended zGNMs.
The data are color-coded by $N$.
$\kappa$ is independent of $t$ for zGNMs of $N\geq5$~MLs, while it is scattered for zGNMs of $N=3$~MLs.
}
\label{fig:kappa-t}
\end{figure}
%%%%%%%%%%%%%%%%%%%%%%%%%%%%%%%%%%%%%%%%%%%%%%%%%%%%%%%%%%%%%

%%%%%%%%%%%%%%%%%%%%%%%%%%%%%%%%%%%%%%%%%%%%%%%%%%%%%%%%%%%%
\begin{figure*}[hbt]
\centering
\includegraphics[width=1.0\linewidth]{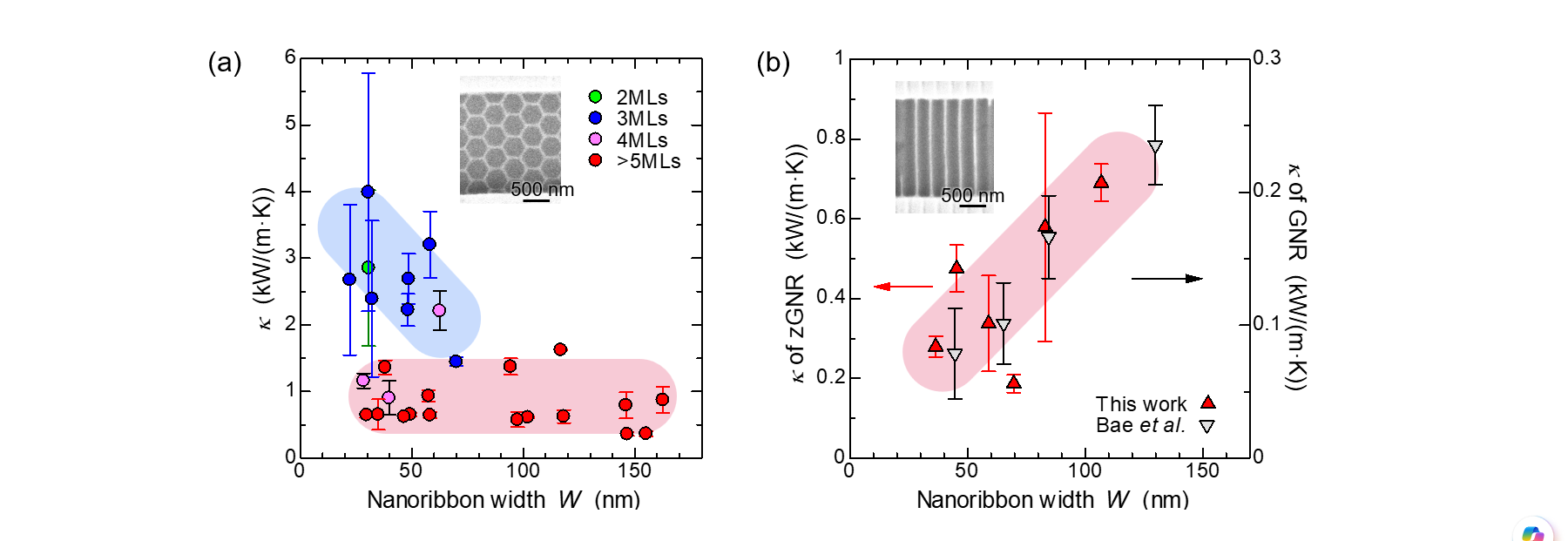}
\caption{
The $W$ dependence of the $\kappa$ of (a) suspended zGNMs and (b) suspended zGNRs ($N\ge5$~MLs).
The $\kappa$ of supported GNR reported in ref.~\cite{Bae2013} is overlaid in (b).
The inset in each figure shows a typical scanning electron microscope image of each device.
The thick blue and red shaded lines show the trend of $W$ dependence of thin and thick devices, respectively.
}
\label{fig:sus_w_all}
\end{figure*}
%%%%%%%%%%%%%%%%%%%%%%%%%%%%%%%%%%%%%%%%%%%%%%%%%%%%%%%%%%%%%

First of all, the $t$ dependence of the $\kappa$ of suspended zGNM is shown in fig.~\ref{fig:kappa-t}.
The overall feature follows that for pristine graphene \cite{Ghosh2010}.
Namely, $\kappa$ decreases with increasing $N$, and saturates for larger $N$.
Similarly for the zGNM in this study, the $\kappa$ is $t$ independent when $N\geq5$~MLs within an error bar.
On the other hand, the $\kappa$ of 3~MLs thick zGNM shows a significantly higher $\kappa$ than that when $N\geq5$~MLs, although it is scattered.
In addition, the $\kappa$ of 2~MLs thick zGNM shows almost the same value as that when $N=3$~MLs, while the $\kappa$ of 4~MLs thick zGNM seems to show an intermediate value between $N\le3$~MLs and $N\ge5$~MLs devices.
Therefore, one can categorize the data into three regions: 2 and 3~MLs thick (thin), 4~MLs thick (intermediate), and $N\ge5$~MLs (thick) regions: it can be considered that the $\kappa$ is $t$ independent in each region.

Figure~\ref{fig:sus_w_all}(a) shows the $W$ dependence of the $\kappa$ of suspended zGNMs.
Unintuitively, the $\kappa$ of thin devices shows a negative trend against $W$, while that of thick devices is almost independent of $W$. 
Importantly, $\kappa$ is not suppressed by nano-structuring for either case.
Since the devices are suspended from the substrate, this unconventional feature is not due to the substrate.
Therefore, it must be a genuine feature of our zGNM.
It can be due to the mesh structure and/or due to the atomically smooth zz edge.

To find the mechanism of this behavior, the $\kappa$ of suspended zGNR was studied in the same manner.
Figure~\ref{fig:sus_w_all}(b) shows the $W$ dependence of the $\kappa$ of suspended zGNR together with that of supported GNR in ref.~\cite{Bae2013}.
Unfortunately, the data is restricted only to the thick device, since it was experimentally difficult to make a sufficient number of thin and suspended zGNRs.
Different from the $\kappa$ of suspended zGNM, the $\kappa$ of suspended zGNR shows intuitive behavior, i.e., $\kappa$ is suppressed by nano-structuring.
It is also consistent with the previous study of GNR~\cite{Bae2013}.
Therefore, from the comparison between zGNM and zGNR, one can conclude that the unconventional $W$ dependence of the $\kappa$ of zGNM is due to the mesh structure rather than the smooth zz edge.

The reason that the mesh structure enhances $\kappa$ is not yet understood.
However, analogous to the PnC~\cite{Zen2014,Anufriev2015}, it is possibly because the wave-like nature of phonons becomes prominent in this nano-scaled periodic mesh structure of graphene, where the phonon mean free path is longer than or comparable to the period of the structure.
As a result, constructive interference of phonons may enhance $\kappa$.
Note that phonons are dominant for thermal transport in graphene and the contribution from electrons can be ignored \cite{Balandin2011, Pop2012}.
Although detailed theoretical consideration is desired to explain this unconventional feature, this may be the first experiment that shows a non-classical feature of phonons even at room temperature.

Apart from this unconventional behavior, it is also found that the $\kappa$ of our zGNR is three to four times higher than that of the GNR~\cite{Bae2013}, even though their $W$ dependence is similar.
This difference can be owed to three differences between devices.
First, the device is thick (5$\sim$10~MLs thick) in this study, while previous study uses monolayer graphene.
According to the $\kappa$ for pristine graphene~\cite{Ghosh2010}, this thickness difference can make the $\kappa$ of the thick zGNR a quarter of that of the monolayer device.
Second, the zGNR is suspended from the substrate, while the GNR~\cite{Bae2013} is supported.
This difference can result in a six to seven times higher $\kappa$ of a suspended device than that of a supported device \cite{Cai2010, Seol2010, Balandin2008}.
Even after considering the two differences above, the $\kappa$ of our zGNR is still about two times higher than that of GNR~\cite{Bae2013}.
And this difference can be due to the third difference between two devices, that is the edge roughness.
The edge of our zGNM is prepared by $\mathrm{H}_2$ plasma etching, and is atomically aligned in the zz structure \cite{Matsui2019,Amend2018}, while atomic-scale order cannot be desired for the edges of GNR~\cite{Bae2013} since it was fabricated by oxygen plasma etching.

This result is consistent with some theoretical studies that predict that $\kappa$ with zz edges can be a few times higher than that with armchair edges reflecting the phonon band difference \cite{Xu2009, Guo2009}, and that $\kappa$ is higher for smooth edges than for rough edges \cite{Evans2010}.
This may be the first experiment that shows the effect of edge atomic structure experimentally.

%%%%%%%%%%%%%%%%%%%%%%
\section{Summary}
%%%%%%%%%%%%%%%%%%%%%%

In this study, zGNM, a regularly patterned graphene mesh with atomically ordered zz edges, suspended from a substrate, was fabricated and its $\kappa$ was measured by the opto-thermal Raman technique.
It was observed that $\kappa$ of thin ($N = 2,3$~MLs) zGNM is inversely proportional to $W$, while that of thick ($N=5$$\sim$10~MLs) zGNM is independent of $W$.
It should be noted that $\kappa$ was not suppressed by nano-structuring even at room temperature for both cases, opposite to the classical picture.
Because the $\kappa$ of suspended zGNR was suppressed by decreasing $W$, it can be concluded that this unintuitive feature is due to the mesh structure.
Probably, constructive interference of phonons enhances $\kappa$ in a periodic mesh structure, the length scale of which is shorter than or comparable to the phonon mean free path.
Further studies such as the dependence on mesh pattern and mesh size are desired to confirm this hypothesis.
Nevertheless, this study can pave the way for the novel features of graphene PnCs.
In addition, the $\kappa$ of zGNR was found to show a higher value than that of GNR with atomically rough edges \cite{Bae2013}.
This is probably due to the effect of atomically ordered zz edges.
This is the first experiment to show the effect of edge atomic structures, which is achieved since a technique to prepare nano-structured graphene with zz edges has finally been established \cite{Yokosawa2022}.


%%%%%%%%%%%%%%%%%%%%%%%%%
\section{Acknowledgement}
%%%%%%%%%%%%%%%%%%%%%%%%%

%\begin{acknowledgement}
%\vspace{1.0 cm}
%\clearpage

A part of this work was supported by ``Nanotechnology Platform Japan'' and ``Advanced Research Infrastructure for Materials and Nanotechnology in Japan'' of the Ministry of Education, Culture, Sports, Science and Technology (MEXT) Grand Number JPMXP09F22UT1064 and JPMXP1223UT1029/JPMXP1224UT1080, respectively, and fabrication was conducted in Takeda Cleanroom with help of Nanofabrication Platform Center of School of Engineering, the University of Tokyo, Japan.

%\end{acknowledgement}

%%%%%%%%%%%%%%%%%%%%%%%%%
%\section{Supplemental}
%%%%%%%%%%%%%%%%%%%%%%%%%

%Figure \ref{fig:G shift} shows the G-band of a zGNM device at different $P_\mathrm{abs}$. 
%It demonstrates that the G-band intensity ($I_\mathrm{G}$) increases, while the G-band frequency ($\omega_\mathrm{G}$) shifts to lower frequency, therefore $T$ increases, as $P_\mathrm{abs}$ increases.

%%%%%%%%%%%%%%%%%%%%%%%%%%%%%%%%%%%%%%%%%%%%%%%%%%%%%%%%%%%%
%\begin{figure}[htb]
%\centering
%\includegraphics[width=0.7\linewidth]{G_shift_v1.png}
%\caption{
%The G-band of a zGNM obtained at various Raman laser power.
%It clearly shows that the $I_\mathrm{G}$ grows, while $\omega_\mathrm{G}$ shifts downwards, as laser power increases.
%}
%\label{fig:G shift}
%\end{figure}
%%%%%%%%%%%%%%%%%%%%%%%%%%%%%%%%%%%%%%%%%%%%%%%%%%%%%%%%%%%%%


%\begin{suppinfo}

%\end{suppinfo}

\bibliography{ThermalCond}

\end{document}